\documentclass[twocolumn,showpacs,showkeys,amsmath,amssymb]{revtex4}
\usepackage{graphicx}% Include figure files
\usepackage{dcolumn}% Align table columns on decimal point
\usepackage{bm}% bold math

\begin{document}

\preprint{APS-v0}

\title{Simulation of Multicellular Tumor Spheroids Growth Dynamics}

\author{Branislav Brutovsky}
\email{bru@seneca.science.upjs.sk}
\author{Denis Horvath}
\email{horvath.denis@gmail.com}
\author{Vladimir Lisy}
\email{lisy@upjs.sk}
\affiliation{Institute of Physics, P.~J.~Safarik University, Jesenna 5, 04154 Kosice, Slovakia}
\date{\today}

\begin{abstract}
The inverse geometric approach to the modeling of the growth of circular
objects revealing required features, such as the velocity of the growth and 
fractal behavior of their contours, is presented. It enables to reproduce
some of the recent findings in morphometry of tumors with the possible
implications for cancer research. The technique is based on cellular
automata paradigm with the transition rules selected by optimization
procedure performed by the genetic algorithms.
\end{abstract}

                             

\maketitle

\section*{Introduction}

Understanding the fundamental laws driving the tumor development
is one of the biggest challenges of contemporary science
\cite{Hanahan2000,Axelrod2006}.
%(Hanahan and Weinberg, 2000; Axelrod et al., 2006)
Internal dynamics of a tumor reveals itself in a number of phenomena,
one of the most obvious ones being the growth. Overtaking its
control would have profound impact to therapeutic strategies.
Cancer research has developed during the past
few decades into a very active scientific field taking on
the concepts from many scientific areas, e. g., statistical physics,
applied mathematics, and nonlinear science
\cite{Baish2000,Delsanto2000,Deisboeck2001,Eloranta1997,Ferreira2002,Chignola2000,
Castorina2006,Galam2001,Gazit1995,Scalerandi2002,Sole2003,Peirolo2004,Spillman2004}.
%(Baish and Jain, 2000; Delsanto et al., 2000; Deisboeck et al., 2001;
%Eloranta, 1997; Ferreira, 2002; Chignola et al., 2000; Castorina et al., 2006;  
%Galam and Radomski, 2001; Gazit et al., 1995; Scalerandi and Peggion, 2002;
%Sole, 2003; Peirolo and Scalerandi, 2004; Spillman et al., 2004)
From a certain point of view, the evolution of tumors 
can be understood as an interplay between the chemical interactions 
and geometric limitations mutually conditioning each other. 
Consequently, it is believed that malignancy of a tumor can be 
inferred exceptionally from the geometric features of its interface 
with the surrounding it tissue 
\cite{Grizzi2005,Escudero2006b}.
%(Grizzi et al., 2005; Escudero, 2006).
Formation of the growing interface is in a continuum approximation 
described by a variety of alternative growth models, such 
as Kardar-Parisi-Zhang
\cite{Kardar1986}
,
%(Kardar et al., 1986),
{\em molecular beam epitaxy} (MBE)
\cite{Krug1997},
%(Krug, 1997),
or Edwards-Wilkinson equations. 
At the same time, the growth 
properties can be classified into 
universality classes 
\cite{Odor2004},
%(Odor, 2004),
each of them showing specific scaling behavior with corresponding 
critical exponents. As implies scaling 
analysis of the 2-dimensional
tumor contours
\cite{Bru1998,Bru2003},
%(Bru et al., 1998, 2003),
the tumor growth dynamics
belongs to the MBE universality class characterized by,
(1)~a linear growth rate (of the radius), (2)~the proliferating activity at the outer border, 
and (3)~diffusion at the tumor surface. 

{\it In vitro} grown tumors usually form 3- (or 2-) dimensional spherical
(or close to) aggregations, called multicellular tumor spheroids (MTS)
\cite{Delsanto2005}.
%(Delsanto et al., 2005). 
These provide, allowing strictly controlled nutritional
and mechanical conditions, excellent experimental patterns to test the validity
of the proposed models of tumor growth
\cite{Preziosi2003}.
%(Preziosi, 2003).
These are usually classified into two groups, 
(1) continuum, formulated through differential equations,
and (2) discrete lattice models, typically represented by cellular automata
\cite{Kansal2000,Patel2001,Quaranta2005},
%(Kansal et al., 2000; Patel et al., 2001; Quaranta et al., 2005),
agent-based
\cite{Mansury2004a},
%(Mansury and Deisboeck, 2004),
and Monte Carlo inspired models
\cite{Ferreira2005a}.
%(Ferreira, 2005). 

Here we present the inverse geometric approach to the MTS growth simulation,
enabling us to evolve an initial MTS by required rate as well as desired
fractal dimension of its contour. The method is based on the cellular automata
paradigm with the transition rules found by the genetic algorithms.

\section*{Simulation and optimization tools}

{\it Cellular automata}~(CA)
\cite{Wolfram1983}
%(Wolfram, 1983)
were originally introduced
by John von Neumann as a possible idealization of biological
systems. In the simple case they consist of a 2D
lattice of cells $s^{(t)}_{ij} \in \{0,1\},\ i, j = -(L_{\rm D}/2),
-(L_{\rm D}/2)+1,\ldots L_{\rm D}/2$, where $t = 0,\ldots, \tau$,
is the time step and $L_{\rm D}$ size of the 2D lattice.
During the $\tau$ time steps they evolve obeying the set of local
transition rules (CA rules) $\sigma$, formally written
\begin{equation}
\begin{array}{ll}
s^{(t+1)}_{ij} = \sigma(& \!\!s^{(t)}_{i-1j-1}, s^{(t)}_{ij-1}, s^{(t)}_{i+1j-1},
s^{(t)}_{i-1j}, s^{(t)}_{ij},\\
& \!\!s^{(t)}_{i+1j}, s^{(t)}_{i-1j+1}, s^{(t)}_{ij+1}, s^{(t)}_{i+1j+1}),
\end{array}
\label{CAmapping}
\end{equation}
which defines the CA rules $\sigma$ as the mapping
\begin{equation}
\sigma: \,
\underbrace{\{0,1\}\times\{0,1\}\times\ldots\times \{0,1\}}_{9}  
\rightarrow \{0,1\}\,.
\label{Mapping}
\end{equation}
Any deterministic CA evolution is represented by the corresponding
point in $2^9$-dimensional binary space enabling, in principle,
immense number of $2^{2^9}$ possible global behaviors, predestining
CA to be the efficient simulation and modeling tool
\cite{Toffoli1987}.
%(Toffoli and Margolous, 1987).
Inherent nonlinearity of CA models is, however, a double-edged
sword. On the one hand, it enables to model a broad variety of
behaviors, from trivial to complex, on the other hand it results
in difficulty with finding the transition rules generating the
desired global behavior.
No well-established universal technique exists to solve the problem,
and, despite sporadic promising applications of genetic algorithms (GA)
to solve the task
\cite{Richards1990,Mitchell1994,Jimenez-Morales2002},
%(Richards et al., 1990; Mitchell et al., 1994; Jimenez-Morales et al., 2002),
one typically implements CA by {\it ad hoc} or microscopically reasoned 
definition of the transition rules.

{\it Genetic Algorithms (GA)}
\cite{Holland1}
%(Holand, 1992)
are general-purpose search and
optimization techniques based on the analogy with Darwinian evolution of biological
species. In this respect, the evolution of a population of individuals is viewed as
a search for an optimum (in general sense) genotype. The feedback comes as 
the interaction of the resulting phenotype with environment. Formalizing 
the basic evolutionary mechanisms, such as mutations, crossing-over and survival
of the fittest, the fundamental theorem of GA was derived
({\it schema theorem}) which shows that the evolution is actually
driven by multiplying and combining {\it good} (quantified by an appropriate
objective function), correlations of traits (also called schemata, or hyperplanes).
The remarkable feature resulting from the schema theorem is the
implicit parallelism stating that by evaluating a (large enough) population of 
individuals, one, at the same time, obtains the information about the quality
of many more correlations of the traits.

Bellow we present the application of GA optimization to find the CA rules
producing the 2D CA evolution by required rate as well
as fractal behavior of the contour.

\section*{Optimization Problem}

In the below numerical studies two competing hypotheses
of the rate of the desired tumor mass production have been
distinguished,

\noindent
a) broadly accepted exponential growth of the tumor mass:
\begin{equation}
M^{(t)} = B+\left[\pi(R^{(0)} )^2-B\right]\exp(C\,t)\,, 
\label{ExponentialGrowth}
\end{equation}
and,

b) the growth of the mass with linearly growing radius
%by Bru et al. (1998, 2003), 
\cite{Bru1998,Bru2003},
\begin{equation}
M^{(t)} = \pi(R^{(0)}+At)^2,
\label{LinearGrowth}
\end{equation}
where $R^{(0)}$ is the initial cluster radius; $A, B, C$ are constants 
parameterizing the growth process. The term $B+\left[\pi( R^{(0)})^2-B\right]$
in (\ref{ExponentialGrowth}) was chosen to start from the initial cluster size
$M^{(0)} = \pi(R^{(0)})^2$ for any $B$.

At the beginning, the chain of concentric circles ({\it patterns})
with randomly deformed close-to-circular contours
$p^{(t)}_{ij} \in \{0,1\},\ i, j = -(L_{\rm D}/2), (-L_{\rm D}/2)+1, \ldots, L_{\rm D}/2$,
for $t = 0,\ldots, \tau$, were generated accordingly to
\begin{equation}
\label{patterns}
p^{(t)}_{ij} =
\left\{
\begin{array}{ll}
0\,, & \mbox{for $\sqrt{i^2+j^2}>R^{(t)}+1$},\\
0\,, & \mbox{or $1$ drawn with probability \small$\frac{1}{2}$}\\
     & \mbox{for $|\sqrt{i^2+j^2}-R^{(t)}|\leq 1$},\\
1\,, & \mbox{for $\sqrt{i^2+j^2}<R^{(t)}-1$},
\end{array}
\right.
\end{equation} 
where the increasing tumor radius $R^{(t)}$ is taken as 
\begin{equation}
R^{(t)} = \sqrt{M^{(t)}/\pi}
\label{RadForExp}
\end{equation}
in the case of exponential tumor mass production Eq.~(\ref{ExponentialGrowth}), and
\begin{equation}
R^{(t)} = R^{(0)}+At
\label{LinRad}
\end{equation}
in the case of linearly growing radius model Eq.~(\ref{LinearGrowth}), respectively.

The optimization task solved by GA was to find the CA transition rules
$\sigma^\ast$ Eq.~(\ref{Mapping}) providing the growth from the initial
pattern $\{s_{ij}^{(0)} \} \equiv \{ p_{ij}^{(0)}\},\ i, j = -(L_{\rm D}/2),
\ldots, (L_{\rm D}/2)-1,L_{\rm D}/2$
through the sequence of the square lattice configurations
$\{s_{ij}^{(t)}\}$, $t = 1,\ldots,\tau$, generated accordingly to
Eq.~(\ref{CAmapping}), with the minimum difference from the above
"prescribed" patterns $\{p_{ij}^{(t)}\}$ in the respective $t$,
quantified by the objective function
\begin{equation}
\label{objective1}
f_1(\sigma)  \equiv  \sqrt{\frac{1}{\tau}\sum^{\tau}_{t=1}
\left(\frac{\sum^{L_{\rm D}}_{i,j} p^{(t)}_{ij}+ 
\sum^{L_{\rm D}}_{i,j} s^{(t)}_{ij}}{1 + w_0 \sum^{L_{\rm D}}_{i,j} 
p^{(t)}_{ij}\delta_{p^{(t)}_{ij} s^{(t)}_{ij}}} 
\right)^2}\,,
\end{equation}
where $\delta$ is the standard Kronecker delta symbol, 
$w_0$ is the weight factor, in our case $w_0 = 2$.
The above expression of the objective function Eq.~(\ref{objective1}) 
reflects the programming issues. The larger overlap 
of $\{s_{ij}^{(t)}\}$ with $\{p_{ij}^{(t)}\}$ enhances the 
denominator of Eq.~(\ref{objective1}), the prefactor $p_{ij}$ 
in the term $p_{ij} \delta_{p_{ij} s_{ij}}$ reduces
the computational overhead by ignoring the calcul
$\delta_{p_{ij} s_{ij}}$ for $p_{ij}=0$.

The other requirement to the desired growth relates to the geometric properties
of the contour/interface. Broadly accepted invariant measure expressing the contour
irregularity is the fractal dimension, $D_{\rm F}$. 
Using the box-counting method
it can be calculated as
\begin{equation}
\label{FractalDimension}
D_{\rm F} = 
\lim_{\epsilon\to 0}\frac{\log N_{\rm B}(\epsilon)}{\log(1/\epsilon)},
\end{equation}
where $N_{\rm B}(\epsilon)$ is the minimum number of boxes of size $\epsilon$
required to cover the contour. Here, it has been determined as the slope in the
log-log plot of $N_{\rm B}(\epsilon)$ against $1/\epsilon$ using the standard
linear fit.

To obtain the CA rules generating the cluster with the required
fractal dimension ($\sigma^\ast$), $D_{\rm F}$, the objective function
Eq.~(\ref{objective1}) has been multiplied by the factor
\begin{equation}
f_2(\sigma) = 1 + w_1 (D^{\tau}_{\rm F}-D_{\rm F})^2,
\label{objective2}
\end{equation}
where $D^{\tau}_{\rm F}$ is the fractal dimension of the cluster kept
after the $\tau $ steps with the CA rules $\sigma$;
the weight $w_1$ was kept 1 in all the presented numerical results.

Finally, the rule-dependent objective function has been written 
\begin{equation}
\label{objective}
f(\sigma) = f_1(\sigma) f_2(\sigma)\,. 
\end{equation}

To sum up, the optimum rule $\sigma^{\ast}$ 
is the subject of GA optimization 
\begin{equation}
\min_\sigma  f(\sigma) =  f(\sigma^{\ast})\,. 
\end{equation}

\section*{Results and discussion}

All the CA runs started from the pattern $\{ p^{(0)}\}$ 
Eq.~(\ref{patterns}), with the radius $R^{(0)} = 5$. 
In all the below GA optimizations, all the CA evolutions ran on the 2-dimensional
box of the size $L_{\rm D}=300$ cells and 
length of the CA evolution $\tau = 100\ (300)$ steps.
The GA search has been applied to find 
the set of CA rules, $\sigma^\ast$,
which gives minimum objective function 
values (Eq.~(\ref{objective1}), or Eq.~(\ref{objective}), respectively).
The size of the population was kept constant (1000 individui),
the probability of bit-flip mutation 0.001, 
the crossing over probability 1,
and ranking selection scheme applied. 
The length of the optimization was $3000$ generations.

\subsection*{Exponential growth vs. linear radius dilemma}

\begin{figure}[h]
\includegraphics[scale=0.34]{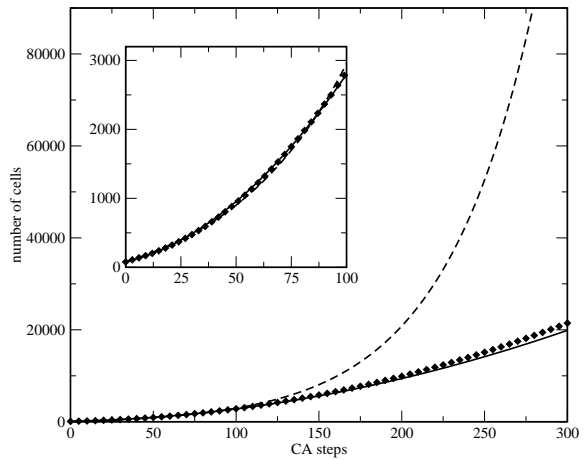}
\caption[]{Comparison of the CA mass production using 
the CA transition rules
obtained by the minimization of Eq.~(\ref{objective1}) during $\tau = 100$ 
steps against the exponential growth Eq.~(\ref{ExponentialGrowth}) (dashed line) 
and the growth with linearly increasing radius Eq.~(\ref{LinearGrowth}) (solid line).
The extrapolation of both the fits beyond the interval of minimization 
is displayed. The inset shows coincidence of the fits on the interval of the minimization.}
\label{shortfit}
\end{figure}

\begin{figure}[h]
\includegraphics[scale=0.34]{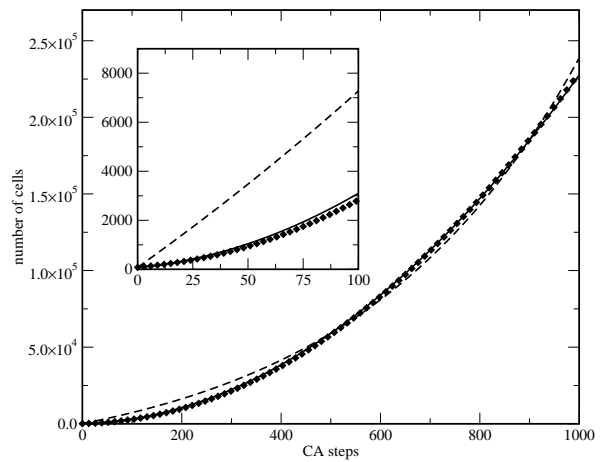}
\caption[]{
Comparison of the CA mass production using the same CA transition rules as in Fig~\ref{shortfit}.
Here, the CA mass production has been fitted by both the fits far beyond the interval of optimization.
The inset shows the comparison of the fits on the interval of optimization.}
\label{longfit}
\end{figure}

The simplest mathematical models of MTS growth
\cite{Shackney1993}
%(Shackney, 1993)
assume exponential increase of the MTS mass during the time 
Eq.~(\ref{ExponentialGrowth}). The above assumption is applied 
mainly because of feasibility of differential and integral calculus, 
nevertheless revealing, hopefully, some of the characteristics of 
the real growth.
Bru et al. \cite{Bru1998,Bru2003}
%Bru et al. (1998, 2003)
have shown from the morphometric 
studies that the mean radius of 2D tumors grows linearly. At the same time, 
they have experimentally shown that the cells proliferation
is located near the outer border of cell colonies. In this work the former
assumption has been tested and compared with the alternative exponential growth
(Figs. \ref{shortfit}, \ref{longfit}). For that reason the GA search has been
carried out to find the CA transition rules which minimize the criterion
Eq.~(\ref{objective1}) with exponentially growing pattern Eq.~(\ref{patterns})
during $\tau = 100$ steps. In the inset of Fig. \ref{shortfit} one can see
that on the interval of the optimization ($\tau = 100$) the CA evolution
can be in principle fitted by the exponential Eq.~(\ref{ExponentialGrowth}) (as it was required)
as well as by Eq.~(\ref{LinearGrowth}), corresponding to the growth with linearly increasing radius,
both with small systematic error, which can be possibly hidden in 
the stochasticity  of the real biological growth. On the other hand, the extrapolation of the fits
beyond the interval of optimization shows obvious divergence of the CA mass
production from the exponential fit, meanwhile its deviation from the regime
with linearly growing radius stays small (nevertheless systematic).
We attribute the latter discrepancy to the fact that the growing interface
during the CA evolution is neither smooth, nor of zero thickness, which 
is true also for real tumors growth.
In Fig.~\ref{longfit} we show 
the comparison of the CA mass production,
using the same CA rules as in Fig.~\ref{shortfit}, fitted by the Eqs.~(\ref{ExponentialGrowth})
and (\ref{LinearGrowth}), respectively, on the
interval much longer than the interval of minimization ($\tau = 100$).
One can see that both the fits are still possible, nevertheless the exponential
fit deviates crucially on the interval of optimization 
(the inset of Fig.~\ref{longfit}).
The above results show better agreement of the CA mass production with
the hypothesis of the MTS growth with linearly growing radius Eq.~(\ref{LinearGrowth}),
with a slight implication towards experiments - the growth of close-to-circular MTS
is probably slightly faster than proposed by Bru et al. \cite{Bru2003},
nevertheless not exponential.

\subsection*{Fractal behavior of the contour}

Figs.~\ref{growth135} and \ref{growth110} show the efficiency of the above approach
to simulate the MTS growth by any required rate (Eq.~(\ref{patterns}) with $R^{(t)}$ coming
from Eq.~({\ref{LinRad})), and, at the same time, desired fractal dimensions of the contour,
$D_{\rm F}$. Here, two GA optimizations have been performed to find the CA rules generating the
mass production minimizing both the criteria Eq.~(\ref{objective1}) and Eq.~(\ref{objective2})
in the multiplicative form Eq.~(\ref{objective}) during $\tau= 300$ steps and
reaching the desired fractal dimensions $D^{\tau}_{\rm F} = 1.35$
(Fig.~\ref{growth135}), and $1.1$ (Fig.~\ref{growth110}), respectively.
The obtained CA rules provide the growth that fits well to the required 
rate as well as desired $D_{\rm F}$.
\begin{figure}[h]
\includegraphics[scale=0.75]{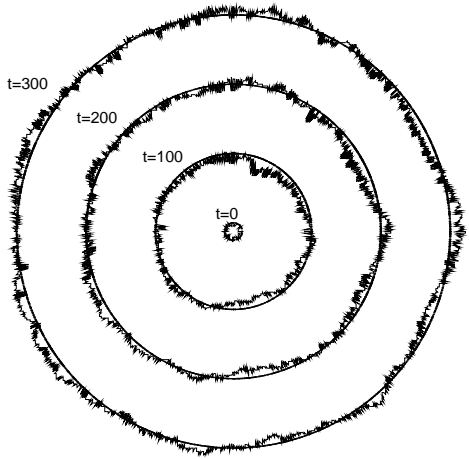}
\caption[]{
The CA mass production using the CA rules obtained by GA minimization of the
Eq.~(\ref{objective}) with the patterns Eq.~(\ref{patterns}) substitute by Eq.~(\ref{LinRad})
with $R^{(0)} = 5$ and velocity constant $A = 0.4$, and the desired fractal dimension
$D_{\rm F}^{300} = 1.35$ (here presented CA run gives $D_{\rm F}^{300} = 1.34)$.
The smooth circles correspond to Eq.~(\ref{LinearGrowth}).}
\label{growth135}
\end{figure}
\begin{figure}[h]
\includegraphics[scale=0.75]{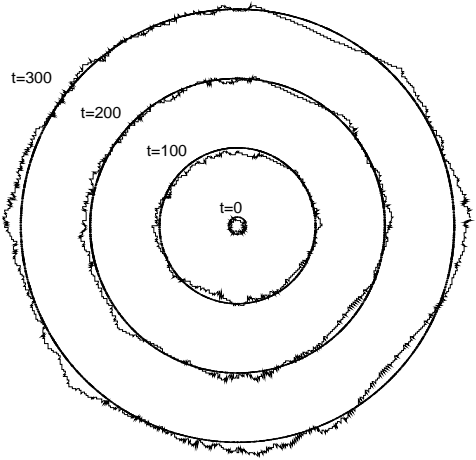}
\caption[]{
The same as in Fig.~\ref{growth135}, requiring the fractal dimension $D_{\rm F}^{300} = 1.1$
(here presented CA run gives $D_{\rm F}^{300} = 1.09)$.}
\label{growth110}
\end{figure}
Fig.~\ref{all100} shows the final size of the CA clusters
using the CA rules obtained by the GA minimization of Eq.~(\ref{objective})
requiring the growth accordingly to Eq.~(\ref{LinearGrowth}) 
during the $\tau = 100$ steps within a broad range of the velocity constants,
$A\in <0.1,0.5>$, as well as the fractal dimensions, $D_{\rm F}^{100}\in<1.0,1.35>$.
Fig.~\ref{a2fd} shows the average $D_{\rm F}^{100}$ generated by
the above CA rules for each of the respective pairs of
the parameters $A$, $D_{\rm F}$. 
\begin{figure}[h]
\includegraphics[scale=0.38]{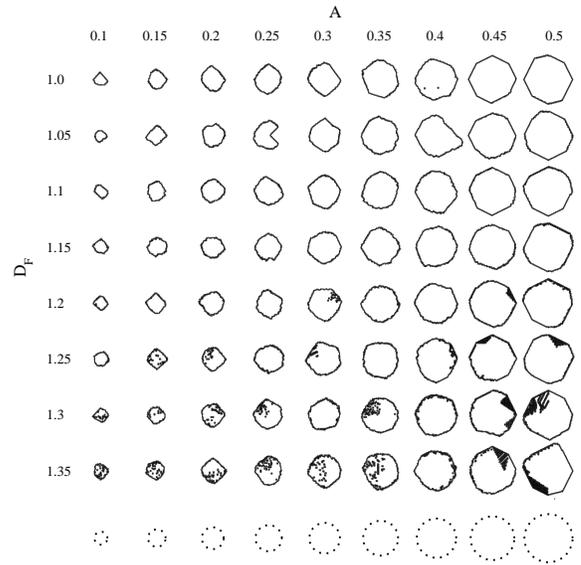}
\caption[]{Resulting contours after $\tau = 100$ steps using the CA rules found  
by the GA for the respective pairs of the desired parameters $A$, $D_{\rm F}$.
The smooth circles correspond to the growth accordingly to Eq.~(\ref{LinearGrowth})
for the respective $A$. The real values of $D_{\rm F}^{100}$ that correspond to the
above CA clusters after $\tau=100$ steps are depicted in Fig. \ref{a2fd}.}
\label{all100}
\end{figure}
\begin{figure}[h]
\includegraphics[scale=0.68]{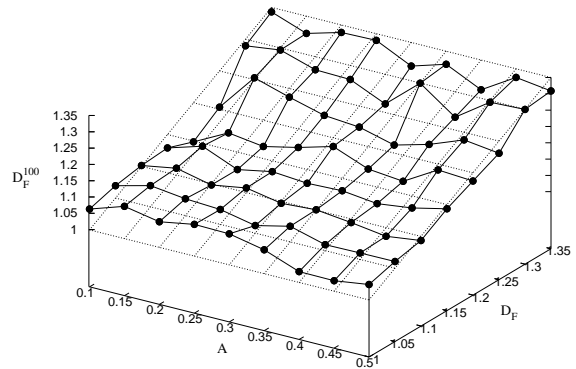}

\vspace{-0.5cm}

\caption[]{
The real fractal dimensions $D_{\rm F}^{100}$ using the CA rules found by GA optimization 
for the respective pairs of $A$, $D_{\rm F}$ (Fig.~\ref{all100}). Dotted plane shows respective 
desired fractal dimensions, $D_{\rm F}$.}
\label{a2fd}
\end{figure}
The above results demonstrate that a specific fractal behavior of the growing 
interface (Fig.~\ref{all100}) out of the broad range ($D_{\rm F}\in <1.05,1.35>$) 
consistent with the morphometric results \cite{Bru1998}
%(Bru et al., 1998)
can be intentionally generated by here presented GA optimization approach. Moreover, 
the growth rate can still be kept at desired value. Beyond these limits 
unwanted artifacts emerge.

\subsection*{Scaling behavior}

%In Refs. \cite{Bru1998,Bru2003} the authors
Bru et al. \cite{Bru1998,Bru2003} used the scaling analysis
to characterize the geometric features of the contours of a few tens
growing tumors and cell colonies. Here we outline the scaling behavior
of the contour of the growing CA cluster resulting from our approach.

A rough interface is usually characterized by the fluctuations
of the height $h(x,t)$ around its mean value $\overline{h}$,
the {\it global interface width}
\cite{Ramasco2000}
%(Ramasco et al., 2000)
\begin{equation}
\label{GlobalInterfaceWidth}
W(L,t) =
\left<\overline{\left[h(x,t)-\overline{h}\right]^2}\right>^{1/2},
\end{equation}
the overbar is the average over all $x$, $L$ is an Euclidean size
and the brackets mean average over different realizations.
In general, the growth processes are expected to follow 
the Family-Vicsek ansatz
\cite{Family1985},
%(Family and Vicsek, 1985)
\begin{equation}
\label{FamilyVicsek}
W(L,t) = t^\beta f(L/\xi(t))\,,
\end{equation}
with the asymptotic behavior of the scaling function
\begin{equation}
\label{guscal1}
f(u) =
\left\{ 
\begin{array}{lll}
u^{\alpha}           &   \mbox{if}  &    u \ll  1\\
\mbox{const}         &   \mbox{if}  &    u \gg  1, 
\end{array}
\right.
\end{equation}
where $\alpha$ is the rougheness exponent and characterizes
the stationary regime in which the height-height correlation length 
$\xi(t)\sim t^{1/z}$ ($z$ is so called dynamic exponent)
has reached a value larger than $L$. The ratio $\beta=\alpha/z$
is called the growth exponent and characterizes the short-time
behavior of the interface. 

To adapt the scaling ansatz to the close-to-circular growing CA cluster
we identify the {\it constant} Euclidean size $L$ with the
{\it time-dependent} mean radius
$\overline{r}\equiv\overline{r}(t)
= (\sum_{k=1}^{ N_{\rm C}(t) } r_k(t))/N_{\rm C}(t)$,
$r_k(t)$ being the distance of the $k$-th contour point from the center 
and $N_{\rm C}(t)$ the numer of contour points in $t$. 
Subsequently, we rewrite Eq.~(\ref{GlobalInterfaceWidth}) into
\begin{equation}
\label{InterfaceWidth}
W_{\rm C}(t) =
\left<\overline{\left[r_k(t)-\overline{r}\right]^2}\right>^{1/2},
\end{equation}
the overbar being the average over all the contour points in $t$
and the brackets mean average over different realizations of contours
reaching the mean radius $\overline{r}$ in $t$, with the scaling ansatz
Eq.~(\ref{FamilyVicsek}) applied
\begin{equation}
\label{FamilyVicsek2}
W_{\rm C}(t) = t^\beta f(\overline{r}/\xi(t)).
\end{equation}
Numerical investigation of the $W_{\rm C}(t)$ (Fig.~\ref{betafit})
reveals the region with the power law behavior. We identify the respective
slope in the log-log plot with the growth exponent $\beta = 0.723$
(fitted in the region $150\leq t\leq 700$).

\begin{figure}[h]
\includegraphics[scale=0.4]{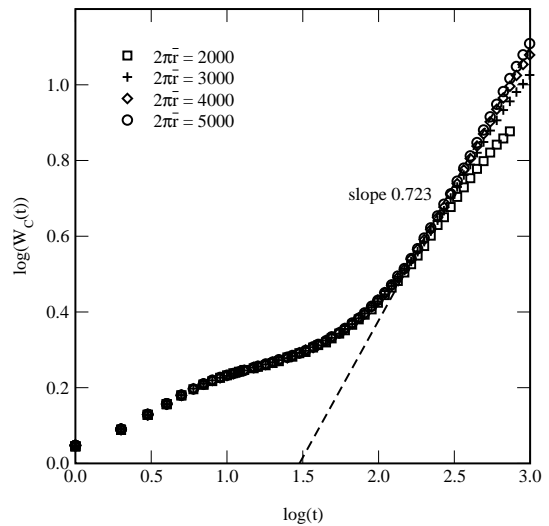}
\caption[]{Scaling behavior of $W_{\rm C}(t)$. The region with 
power law behavior can be identified, corresponding to the slope $0.723$.} 
\label{betafit}
\end{figure}
To draw more complete scenario of the growing CA cluster scaling
behavior, the deeper investigation of site-site correlation
functions  in both radial and poloidal directions is needed.
This type of studies will follow.   

\section*{Conclusion}

In the paper we have presented the approach to the modeling of multicellular tumor spheroids 
by required growth rate and fractal dimension. The technique is based on the combination of the CA 
modeling with the transition rules searched by the GA.  Here demonstrated results show the feasibility 
of the approach in this specific case. Based on the similarity of the geometric properties 
of the CA evolution and the tumor contours (such as locality of the interaction/communication, 
deformed contour and nonzero thickness of the proliferating layer) we have reasoned that the
MTS mass production is slightly faster than corresponding to linearly growing radius \cite{Bru2003}.
%(Bru et al., 2003).
At the same time our results imply that the often used 
Gompertz curve of the tumor mass progression comes as a higher level phenomena related to the nutrition, 
space restrictions, etc. We believe that our approach could be implemented as the backbone 
into the more sophisticated models of tumor growth encompassing the above higher-level 
mechanisms.
Computationally efficient {\it on the fly} scaling analysis during
the CA evolution would enable to bias the GA optimization towards
the MTS growth with desired scaling properties of the contour;
nevertheless its efficient realization is the subject of our ongoing
research. Successful implementation of scaling analysis into the
optimization process could significantly contribute to the discussion
on the scaling behavior of the real tumors
\cite{Bru2003,Buceta2005,Bru2005}.
%(Bru et al., 2003, 2005; Buceta and Galeano, 2005).

In our opinion the above presented optimization approach to the modeling
of growing clusters by required rate and surface properties can find applications 
in many different fields, such as molecular science, surface design 
or bioinformatics.

\begin{acknowledgments}
A part of this work has been performed under the Project HPC-EUROPA (RII3-CT-2003-506079),
with the support of the European Community - Research Infrastructure Action under
the FP6 "Structuring the European Research Area" Programme.
\end{acknowledgments}

%\bibliography{cancer,gapub}

\end{document}